\documentclass[aps,pra,showpacs,twocolumn,nofootinbib]{revtex4-1}
\usepackage{amsmath,graphicx,color,amssymb,microtype}

\newcommand{\pr}[1]{\mathbb{P}(#1)}
\newcommand{\condpr}[2]{\pr{#1 | #2}}

\newcommand{\theo}[1]{\begin{quote}{\it #1}\end{quote}}

\usepackage{hyperref}

\begin{document}

\title{Bell's Theorem, Accountability
and Nonlocality}

\author{Nicola Vona}
\email{vona@math.lmu.de}
\affiliation{Mathematisches Institut der LMU M\"unchen, Theresienstra\ss e 39, D-80333 M\"unchen, Germany}

\author{Yeong-Cherng~Liang}
\email{yliang@phys.ethz.ch}
\affiliation{Institute for Theoretical Physics, ETH Zurich, 8093 Zurich, Switzerland}
\affiliation{Group of Applied Physics, University of Geneva, CH-1211 Geneva, Switzerland}

\date{\today}

\begin{abstract}
Bell's theorem is a fundamental theorem in physics concerning the incompatibility between some correlations predicted by  quantum theory and a large class of physical theories. In this paper, we introduce the hypothesis of accountability, which demands that it is possible to explain the correlations of the data collected in many runs of a Bell experiment in terms of what happens in each single run. 
Under this assumption, and making use of a recent result by Colbeck and Renner [Nat. Commun. {\bf 2}, 411 (2011)], we then show that  any nontrivial account of these correlations in the form of an extension of quantum theory must violate parameter independence.
Moreover, we analyze the violation of outcome independence of quantum mechanics and show that it is {also } a manifestation of nonlocality.
\end{abstract}

\maketitle

\section{Introduction}
Beyond {the} shadow of a doubt, quantum theory is one of the most successful theories to date, showing exceptional empirical adequacy in numerous experiments. 
The very operational nature of the standard theory, however, has caused much controversy on what the theory actually says about the world  beyond empirical predictions. 
For instance, such controversy has led to the celebrated paper by Einstein-Podolsky-Rosen (EPR) \cite{EPR35}, questioning the completeness of quantum theory as a physical theory.

For almost 30 years, the paradoxical situation raised by EPR, and the possibility to complete quantum theory in the form of a local hidden-variable theory received almost no attention among the  physics community,  presumably considering the question metaphysical. It thus came as a surprise when Bell showed that the empirical adequacy of all local hidden-variable theories, regardless of their details, is actually {\it  falsifiable} against the predictions of quantum theory \cite{Bell1964}.
 
More precisely, Bell showed that the correlations of measurement outcomes achievable in any {\it  Bell-local} theory---with local hidden-variable theories being an example---must satisfy some constraints in the form of inequalities, now commonly referred to as Bell { inequalities}.
Moreover, there are probability distributions of measurement outcomes predicted by quantum mechanics that violate such inequalities.

Although some loopholes were always present, many independent experiments showed a clear violation of Bell inequalities,\footnote{See, for instance, Refs.~\cite{Aspect:1999,GiustinaMechRamelow2013,ChristensenMcCuskerAltepeter2013} and references therein; 
a discussion of the various loopholes in Bell experiments can be found in the review paper~\cite{Brunner:2013}.} 
 most notably in the form due to Clauser, Horner, Shimony, and Holt (CHSH) \cite{ClauserHorneShimony1969}. 
Therefore, {\it there are experimental results that cannot be accounted for by any Bell-local theory}. Despite that, the implications of these empirical observations and {of} Bell's theorem itself are still highly debated (see, for instance, Refs.~\cite{ClauserShimony1978,Jarrett1984,Griffiths1987,Shimony,Mermin,Dickson1998,BertlmannZeilinger2002,Bell2004,Norsen2009,SeevinckUffink2011,Gisin2012,CavalcantiWiseman2012,Henson2013,Berge} and references therein).
The importance of the theorem lies in its generality, but this feature makes it also prone to different interpretations.

{In this work we put forward the hypothesis that the correlations that appear after many  runs of a Bell-like experiment are the consequence of a process that takes places at every run of the experiment.
Then, using a recent result of Colbeck and Renner~\cite{ColbeckRenner2011}, we show that any theory that accounts for such a process and is in accordance with quantum mechanics must violate parameter independence~\cite{Jarrett1984,Shimony}.
The connection of this nonlocality with the violation of outcome independence of quantum mechanics will also be discussed.}

The rest of this paper is organized as follows. We will begin by { recalling } Bell's theorem in Sec.~\ref{Sec:BellTheorem}.  Then, we give an exposition of the Bell-locality condition and some immediate implications of Bell's theorem in Sec.~\ref{Sec:BellLocal+Implication}. After that, in Sec.~\ref{Sec:Sharpen} we introduce the hypothesis of accountability and present our main results. We conclude by making some further remarks on the hypothesis of accountability and providing a brief summary in Sec.~\ref{Sec:Conclusion}.

\section{Bell's Theorem}\label{Sec:BellTheorem}

\begin{figure}[h!]
\centering \includegraphics[width=.95\columnwidth]{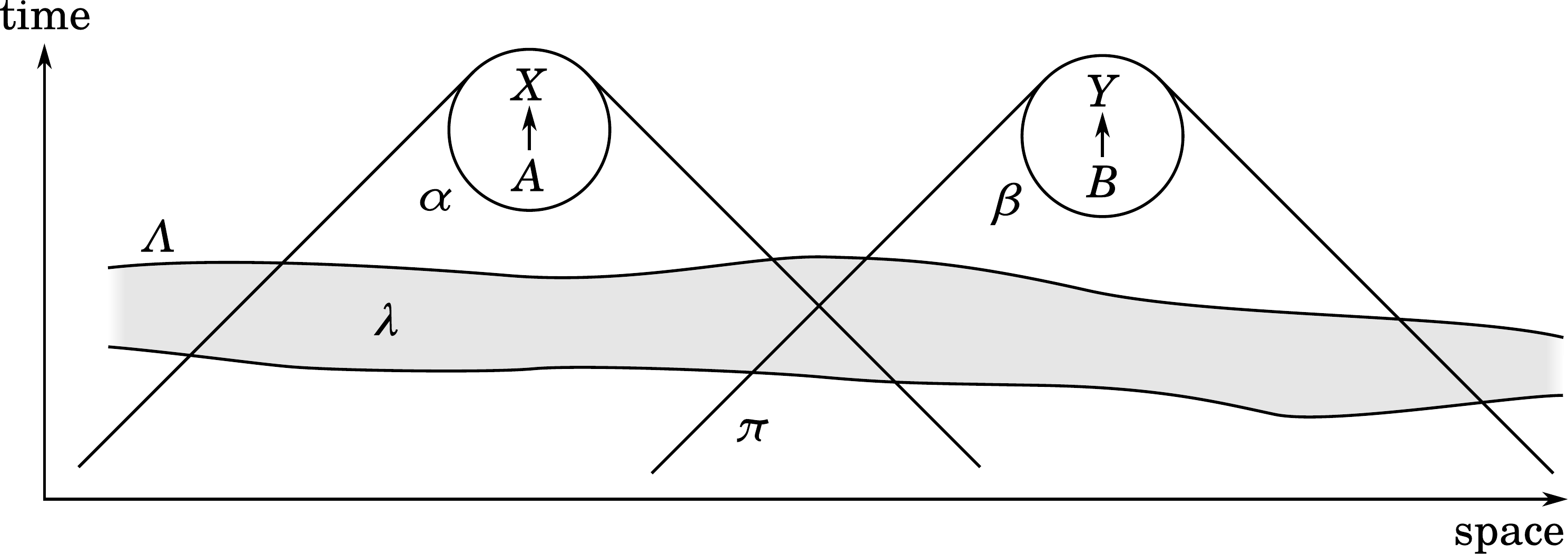}  
\caption{Spacetime diagram of a Bell experiment. The spacetime region $\alpha$ encloses $A$ and $X$, while the region $\beta$ encloses $B$ and $Y$.  The spacetime region $\Lambda$ (shaded), to which the information $\lambda$ pertains, shields $\alpha$ and $\beta$ from their common past $\pi$.}
\label{fig:BellRegions}
\end{figure}

Let us now briefly recall Bell's theorem, stating it in the notation of  Ref.~\cite{ColbeckRenner2011} (see also Fig.~\ref{fig:BellRegions}).
Consider two physical systems, generated by a common source, and moving in opposite directions.
The first one enters a measurement apparatus which has a number of possible settings, denoted by the parameter $A$. 
The spacetime region where the choice of the value $A$ and the measurement\footnote{By measurement, we mean the physical process by which an outcome is produced upon setting the measurement apparatus to the state $A$.} are performed is denoted by $\alpha$, and the result of the measurement by $X$.
Similarly, the second physical system enters an apparatus with setting $B$ and prompts the outcome $Y$.
The choice of the value $B$ and the {corresponding} measurement are performed in the spacetime region $\beta$.

Given a theory, we denote by $\lambda$ the {\it  complete} specification, according to that theory, of all information {relevant to make predictions for the experiments performed in $\alpha$ and $\beta$ and} pertaining to a spacetime region {$\Lambda$} like that depicted in Fig.~\ref{fig:BellRegions}.
{In other words, given $\lambda$, any further information about the two systems in $\Lambda$ and in its past  is redundant for the purpose of making predictions with the theory under consideration.\footnote{For a more detailed discussion, see, for instance, Chapter 24 of Ref.~\cite{Bell2004} and Refs.~\cite{Bricmont1999,Norsen2009,SeevinckUffink2011,Norsen2011}.}}
For example, $\lambda$ can include the specification {of the joint state of the two systems and} of the experimental set-up used to prepare it, but also any other kind of physical object (classical configurations, quantum wave functions, classical and quantum fields, random variables, etc.).
Note that our definition of $\lambda$ differs from that of Bell, who considers only local beables~\cite{Bell2004}, but also from that of Colbeck and Renner~\cite{ColbeckRenner2011}, who allow for a bigger spacetime region.

Following Bell (see Chapter 12 of Ref.~\cite{Bell2004}), we assume throughout
that the choice of a particular value for the parameter $A$ can be made freely inside $\alpha$, i.e.\ independently of the choice of the parameter $B$ and of the state $\lambda$; a corresponding assumption holds of course for $B$.
This means
\begin{equation}\label{eq:NoConspiracy}
\begin{aligned}
\condpr{A}{B,\lambda} &= \pr{A},
\\
\condpr{B}{A,\lambda} &= \pr{B},
\end{aligned}\end{equation}
where $\condpr{l}{m}$ denotes the probability of $l$  given $m$. 
We refer to this assumption as {\it no-conspiracy}. 
Moreover, a theory is said to be {\it Bell-local} if 
\begin{equation}\label{eq:BellLocality}
\condpr{X,Y}{A,B,\lambda}
=\condpr{X}{A,\lambda} \ \condpr{Y}{B,\lambda}  ,
\end{equation}
that is equivalent to\footnote{See Appendix~\ref{App:BellLocal} for a proof of the equivalence.}
\begin{equation}\label{eq:BellLocality2}
\left\{ \begin{aligned} 
&  \condpr{X}{Y,A,B,\lambda} = \condpr{X}{A,\lambda}, 
\\
&\condpr{Y}{X,A,B,\lambda} = \condpr{Y}{B,\lambda},
  \end{aligned} \right.   
\end{equation}
i.e., once $\lambda$ is specified, the probabilities of the outcomes given their local settings become statistically independent. 
From the above conditions one can prove that: 
\theo{If a Bell-local theory that describes the outcomes of this experiment exists, then the CHSH-Bell inequalities~\cite{ClauserHorneShimony1969} hold.}
Nevertheless, appropriate measurements on the singlet state {of a pair of entangled spin-$1/2$ particles} give rise to correlations of measurement outcomes that violate Bell inequalities, and thus~\cite{Bell1964}:
\theo{No Bell-local theory is compatible with all quantum mechanical predictions.}

\section{{The} Bell-Locality Condition and The Implications of Bell's Theorem}
\label{Sec:BellLocal+Implication}

As mentioned above, Bell inequalities have been experimentally violated in various physical systems. Assuming no-conspiracy,  and that the loopholes in the performed experiments are not responsible for the observed Bell inequality violation, we get that 
\theo{no Bell-local theory is  compatible with all existing empirical observations,} 
a matter of fact that is arguably more profound than the incompatibility between quantum theory and any Bell-local  theory.
This empirical fact will stay regardless of what our best physical theory will be in the future.

A source of many difficulties in appreciating Bell's theorem is the fact that, 
using a certain theory to describe the experiment,
 an {\it apparent} violation of  the Bell-locality  condition, Eq.~\eqref{eq:BellLocality}, can also happen due to 
the incompleteness of the theory at hand, a fact discussed at length by Bell himself (see for example Chapters 16 and 24 of Ref.~\cite{Bell2004}).

To understand the implications of Bell's theorem, let us start by getting a better insight into the meaning of Bell-locality. To this end, we note that a very influential account of the meaning of Bell-locality was given by Jarrett~\cite{Jarrett1984}, who proposed to make {Eq. }\eqref{eq:BellLocality} more explicit by substituting it with two conditions: {\it parameter independence}
\begin{equation}\label{eq:ParameterIndependence}
\begin{aligned}
\condpr{X}{A,B,\lambda} &= \condpr{X}{A,\lambda},
\\
\condpr{Y}{A,B,\lambda} &= \condpr{Y}{B,\lambda},
\end{aligned}
\end{equation}
and {\it outcome independence}\footnote{The names for these conditions were introduced by A.~Shimony in Ref.~\cite{Shimony}.}
\begin{equation}\label{eq:OutcomeIndependence}
\begin{aligned}
\condpr{X}{Y,A,B,\lambda} &= \condpr{X}{A,B,\lambda},
\\
\condpr{Y}{X,A,B,\lambda} &= \condpr{Y}{A,B,\lambda}.
\end{aligned}
\end{equation}
As a consequence of the identity
\begin{equation}\label{eq:Id}
\condpr{X,Y}{A,B,\lambda} 
	= \condpr{X}{Y,A,B,\lambda}\,  \condpr{Y}{A,B,\lambda}   ,
\end{equation}
the  requirement of Bell-locality is  equivalent to the conjunction of these conditions~\cite{Jarrett1984}:
\begin{equation}
\text{\it {Bell-locality}} 
\Leftrightarrow
\left\{ \begin{aligned} 
&\text{\it parameter independence}   
\\
&\text{\it outcome independence.  } \end{aligned} \right.
\end{equation}
A violation of Bell-locality is thus a violation of at least one of the two independence conditions mentioned above.

The relation between parameter independence and locality is evident: if turning the knob that controls the setting $A$, one is able to affect  what happens in the region $\beta$, although $\alpha$ and $\beta$ are spacelike-separated, then a nonlocal influence is clearly present.\footnote{Here {\it nonlocal influence} does not necessarily refer to the presence of a superluminal entity as in Ref.~\cite{Bancal2012,Barnea2013}, but to any possible {\it physical} relationship between the two events.} Note that a violation of parameter independence alone does not imply that it is practically possible to exchange superluminal signals between the regions $\alpha$ and $\beta$, for which one needs also complete control on  $\lambda$, or that the violation of Eq.~\eqref{eq:ParameterIndependence} is still present after  averaging over all possible values of $\lambda$ compatible with the preparation procedure.
 This averaged version of Eq.~\eqref{eq:ParameterIndependence} reads
\begin{equation}\label{eq:no-signaling}
\begin{split}
{\condpr{X}{A,B}} 
&=
\int  \condpr{X}{A,B,\lambda} \, {\pr{\lambda} \,}{\rm d}\lambda 
\\
&= \int \condpr{X}{A,\lambda}\,{\pr{\lambda} \,}{\rm d}\lambda
=\condpr{X}{A},
\\
{\condpr{Y}{A,B}} 
&=
\int \condpr{Y}{A,B,\lambda}\,{\pr{\lambda}\,}{\rm d}\lambda 
\\
&= \int \condpr{Y}{B,\lambda}\,{\pr{\lambda} \,}{\rm d}\lambda
=\condpr{Y}{B},
\end{split}
\end{equation}
where $\pr{\lambda}$ is the probability of $\lambda$ for a given preparation procedure. 
Equation~\eqref{eq:no-signaling} is referred to as the {\it signal locality} condition in Ref.~\cite{CavalcantiWiseman2012}, as the {\it relativistic causality} condition  in Ref.~\cite{PopescuRohrlich:1994}, and {as} the {\it no-signaling}  condition  in Ref.~\cite{Barrett:2005}, and is a weaker requirement than Eq.~\eqref{eq:ParameterIndependence}. 

The meaning of outcome independence is more intricate, as  its violation can have   different origins.
The first one is the {\it incomplete} specification of the common past of the two subsystems. 
Consider for example a theory  such that the outcomes $X$ and $Y$ are completely determined by the initial state $\lambda$, and therefore any influence between events in $\alpha$ and $\beta$ can be excluded beyond any doubt.
Suppose for the sake of argument that we may split $\lambda$ into two parts $\lambda_0$ and $\lambda_1$, and consider a new description of the experiment in which $\lambda_1$ is ignored. The outcome $X$ in $\alpha$ will then conceivably disclose some information about $Y$---missing in the description of the state $\lambda_0$ and useful to predict the outcome $Y$---that creates a correlation between $X$ and $Y$.
Such a violation of outcome independence is due to the incompleteness of the description $\lambda_0$ of the state of the joint system in the theory at hand, and can be overcome by considering an extended theory that recovers the missing information.

However, a more fundamental violation of outcome independence is possible too.
The determination of a particular value $x$ for the outcome $X$ is an event that happens in the spacetime region $\alpha$, as well as the determination of a value $y$ for $Y$ is an event that happens in the region $\beta$.
These events could have an impact {on each other}---through some presently unknown physical mechanism---directly and independently of the past  state $\lambda$, or they could both depend on some other event  lying {outside} the common past of $\alpha$ and $\beta$.
Therefore, also a violation of outcome independence can be the signature of a nonlocal influence.

Bell-locality, parameter independence, and outcome independence are all notions relative to one specific theory: they all involve the description $\lambda$ of the state of the system, that is clearly different for different theories.
As a consequence, {depending on the theory,} the same phenomenon may have some account that is Bell-local [i.e. satisfying Eq.~\eqref{eq:BellLocality}] and some other that is not, some that fulfills parameter independence and some that does not, some that  fulfills outcome independence and some that does not, independently of the actual presence of nonlocal influences.
The problem is always the interplay of {\it locality} and {\it completeness}.
Bell's theorem is of such great importance precisely because it is a statement that applies to any possible theory that accounts for the quantum violation, and more generally for the experimentally observed violation of  Bell inequalities: {\it none} of them is Bell-local. Bell and some of his commentators concluded from this that the incompleteness of a specific theory---something that can be eliminated {by} considering an extended theory---cannot be the exclusive cause of the correlations found in the outcomes, and a genuine nonlocal influence must be present.
This is however not the unanimous view among  physicists, as can be seen  in the following passage by Dickson (from page 138 of Ref.~\cite{Dickson1998}):
\begin{quote}
Some authors have suggested that the violation of outcome independence rather than parameter independence saves a theory from superluminal causation, and should be understood instead as a consequence of the fact that the particles are not ontologically distinct (despite their spatio-temporal separation). 
[\dots]
Standard quantum mechanics might, for example, make plausible the idea that the particles are not distinct individuals, because it does not assign either of them its own statevector.
\end{quote}

\section{The hypothesis of accountability and  the Implications of Bell's Theorem}\label{Sec:Sharpen}

{A working principle that is implicitly, but widely adopted by physicists is that any observed phenomenon is accountable in terms of accepted facts and fundamental principles of Nature.}

{In  the described Bell  experiment the outcomes $X$ and $Y$ are correlated, but the initial quantum state of the composite system is not sufficient to predict their actual values in any given run of the experiment: quantum mechanics does not explain the process that instantiates the outcomes in each single run.
The values of $X$ and $Y$ are such that repeating the experiment many times a correlation between them shows up.
The production of the outcome $X$ is not completely (statistically) independent from the event producing the outcome $Y$: when $X$ is determined, $Y$ must somehow be taken into account in such a way to build up the correlations run after run.
This is an actual phenomenon, and as with any observed phenomenon it demands an explanation; we give a formal standing to this demand by the following \emph{hypothesis of accountability}:
\theo{
The correlations between the measurement outcomes $X$ and $Y$ are the consequence of an accountable, but not necessarily deterministic process that takes place at each single run of the experiment.} }

\subsection{All nontrivial extensions of quantum theory violate parameter independence}

{If we maintain that quantum mechanics delivers the correct probability distributions for measurement outcomes, then assuming the hypothesis of accountability is equivalent to assuming  that a theory exists, whose predictions are more refined than those of quantum mechanics yet being always in accordance with them.}
For such a theory
\begin{equation}\label{eq:HV}
	\lambda=(\psi,\Xi), 
\end{equation}
where $\psi$ is the usual quantum-mechanical wave function  of the composite system, and $\Xi$ is some additional structure such that
\begin{equation}
	\int \condpr{X,Y}{A,B,\psi,\Xi}\, \condpr{\Xi}{\psi} \, {\rm d}\Xi  =  \condpr{X,Y}{A,B,\psi} .
\end{equation}
The authors of Ref.~\cite{ColbeckRenner2011} considered a theory of this kind, with no specific assumption on the nature of $\Xi$ except  that it is accessible from any spacetime point and that it is static, i.e.\ its behavior does not depend on where and when it is observed.
Then, they showed that the so-called FR condition,%
\footnote
{The FR condition was introduced by the authors of Ref.~\cite{ColbeckRenner2011} to capture the assumption that the measurement settings can be chosen freely, as an alternative to Eq.~\eqref{eq:NoConspiracy}. There is currently no unanimous agreement on which of these conditions represents the right formulation of this assumption. For a discussion on this, see for instance Refs.~\cite{GhirardiRomano2013,GhirardiRomano2013a} and Ref.~\cite{ColbeckRenner2013}.}
 namely
\begin{equation}\label{eq:FreeChoice}
\begin{split}
&\condpr{A}{Y,B,\lambda} = \pr{A},
\\
&\condpr{B}{X,A,\lambda} = \pr{B},
\end{split}\end{equation}
implies that 
\begin{equation}\label{eq:NoExt}
\condpr{X,Y}{A,B,\psi,\Xi} = \condpr{X,Y}{A,B,\psi},
\end{equation}
i.e.\ 
{\it the additional information provided by any  extended theory {satisfying Eq.~\eqref{eq:FreeChoice}} does not refine the predictions of quantum mechanics about the outcomes $X$ and $Y$}. Given the initial quantum state $\psi$ of the composite system, the values of the parameters $A$ and $B$ represent the only information relevant  to predict the outcomes $X$ and $Y$, any further information being redundant.

It is important to note that the FR condition,  Eq.~\eqref{eq:FreeChoice}, is equivalent to the conjunction of the conditions of no-conspiracy, Eq.~\eqref{eq:NoConspiracy}, and of parameter independence, Eq.~\eqref{eq:ParameterIndependence} (see Appendix~\ref{App:FR} for a proof). In other words, the result of Ref.~\cite{ColbeckRenner2011} shows that
\begin{equation}
	\text{Eq.~\eqref{eq:HV}} \land \text{Eq.~\eqref{eq:NoConspiracy}} \land \text{Eq.~\eqref{eq:ParameterIndependence}} 
	\Rightarrow \text{Eq.~\eqref{eq:NoExt}} {.}
\end{equation}
Thus, assuming no-conspiracy, any physical theory satisfying Eq.~\eqref{eq:HV} and providing  a more refined prediction for  the measurement outcomes  must necessarily violate  parameter independence, a feature that is indisputably nonlocal.
{{\it The validity of quantum mechanics together with the hypothesis of accountability inevitably lead to nonlocality.}}

\subsection{Nonlocality from the violation of outcome independence}

{The proof of the previous section does not explain how the violation of outcome independence of quantum mechanics is related to the nonlocality of the extended theory.
We now analyze this connection.}

{The hypothesis of accountability together with the fact that the quantum state does not describe the process that instantiates the {\em correlated}  outcomes at each run, imply that the actual values of the measurement outcomes $X$ and $Y$ are determined in the regions $\alpha$ and $\beta$ taking into account some further information besides the quantum state.
Then, at least one of the following must be true:}
\begin{enumerate}
\item 
{$X$ and $Y$ depend} at least partially on some event  outside the common past $\pi$ of $\alpha$ and $\beta$;\label{item:OutsidePast}
\item 
{$X$ and $Y$ depend} at least partially on some process that  ``lives'' outside of spacetime, i.e.\ that takes place in a space different from spacetime, as for instance a space of parameters, labels, or outcomes;\label{item:OutsideSpaceTime}
\item 
{$X$ and $Y$ depend} exclusively on events that happened inside the common past $\pi$.\label{item:InsidePast}
\end{enumerate}

In the first case (for instance when the outcomes  are entirely decided inside the regions $\alpha$ and $\beta$), the correlations are explained through a  nonlocal influence directly intervening in the process that decides the outcomes.\footnote{Since joint conditional quantum probabilities do not depend on time ordering of local measurements, such an influence can not correspond to any physical entity traveling in spacetime with finite speed.}

In the second case there is no direct nonlocal influence between spacelike-separated events; nevertheless, there is a process not localizable within spacetime that is able to influence the outcomes.
Since the measurement events are spacelike-separated, so this process is able to ``transcend'' spacetime and to act nonlocally.

{The last case is the only one in which the violation of outcome independence is not manifestly connected to nonlocality.
However, it is not correct to think that in this case outcome independence is violated  simply because the quantum state misses some event in $\pi$.
To appreciate this point, consider a theory in which the quantum state is supplemented by the specification of these events.
If the problem was solely due to missing information, then this extended theory would satisfy outcome independence, and therefore also Bell-locality,  but this is ruled out by Bell's theorem and by the observed violation of Bell inequalities.
Hence, the extended theory still needs to violate outcome independence or parameter independence.
Recall that  the extended theory has been introduced to account for the correlations, therefore it must be able to deliver predictions more refined than those of quantum mechanics.
The result of Ref.~\cite{ColbeckRenner2011} implies that it is {\em impossible}  to refine the predictions of quantum mechanics by supplementing the quantum state with whatever additional information without violating parameter independence (maintaing always that no-conspiracy holds).
Therefore, the extended theory must violate parameter independence and so being manifestly nonlocal.
}

Summarizing,  the hypothesis of accountability and the result of Ref.~\cite{ColbeckRenner2011} imply that {the only cases in which there can be} an explanation that does not violate parameter independence 
are case~\ref{item:OutsidePast} and~\ref{item:OutsideSpaceTime} from the previous list.
Therefore, some nonlocal feature is always present.

Let  us remark that a  theory for which case~\ref{item:OutsidePast} or~\ref{item:InsidePast} are verified is easy to recognize; not so for a theory realizing case~\ref{item:OutsideSpaceTime}.
An example of the latter would be an extension of quantum theory that consists in taking literally   the many worlds interpretation~\citep{DewittGraham1973}.
To see this, consider a single world among the many; when a measurement is performed, the considered world has to ``decide'' which outcome to present.
The precise way in which this process works is not known, and one can imagine it to take place in the space of the possible outcomes, worlds, or in some  other space, but clearly it is {\em not} a process that can be accounted for in the spacetime that we are familiar with. 
In our setting, when the measurement is performed in the regions  $\alpha$ and $\beta$, the pointers of the respective apparatuses have to  ``get in touch'' with the mentioned process that decides the outcomes in the considered world.
Therefore, the process of decision must be able to somehow influence both the regions $\alpha$ and $\beta$, exerting on them a nonlocal action.
{The same is true for consistent histories \citep{Griffiths1987}, provided the considered world is substituted by the actual history that is realized by a given run of the experiment.}
An analogous argument applies also to the interpretation according to which quantum mechanics is local because the two outcomes do not make sense separately, but only as a couple, generated together~(cf.\ passage by Dickson quoted at the end of Section~\ref{Sec:BellLocal+Implication}). Indeed, even if the two outcomes are generated together in some outcomes space, outside spacetime, they must still  somehow be delivered to the regions $\alpha$ and $\beta$.

{Of course, it is conceivable that future physical theories do not satisfy Eq.~\eqref{eq:HV}. 
For instance, the analog of Eq.~\eqref{eq:HV} for Newtonian and quantum mechanics would be $\psi = (q,p,\Xi)$, where $q$ are positions and $p$ momenta, that is clearly not the case.
On the contrary, it is likely that future theories will reduce to quantum mechanics only in an approximate sense or only in some limit.
In this case we can not make direct use of the result of Ref.~\cite{ColbeckRenner2011}.
Nevertheless, it is a fact that quantum mechanics is in very good agreement with the experimental results so far collected in Bell-like settings.
As a consequence, any future theory, however different from quantum mechanics it may seem , must give predictions for Bell experiments very close to those of quantum mechanics; restricting the new theory to this specific kind of set-up, it is then possible to repeat our arguments.
}

\section{Concluding remarks}
\label{Sec:Conclusion}

The validity of the hypothesis of accountability is fundamental for our general argument, yet it is not universally accepted among the physics community.
Some researcher maintains the view that correlations simply happen,\footnote{For instance, the slogan of ``correlations without correlata'' was proposed by N. D. Mermin in Ref.~\cite{Mermin}.} 
and that no further account beyond the quantum measurement process is possible for the mechanism that gives rise to these correlations at every single run (see, for instance, Ref.~\cite{Berge}).
This position can not be excluded on a purely logical basis, but it is clearly against the spirit of physics itself, that constantly seeks for a deeper insight into the underlying cause of any observed phenomenon.
Regarding a phenomenon as non-accountable, in the authors' opinion,  represents a dangerous move in any scientific endeavor. If an observed phenomenon does not have an explanation in terms of accepted facts and principles of Nature, it should be taken as a sign that some of these premises  have to be revised. For sure, the progress of science would be extremely limited if renouncing  the explicability of unexpected phenomenon were accepted as part of the scientific methods.

To summarize, under the assumptions that (1) measurement settings can be chosen freely independent of each other and of the prior state of the physical system to be tested, and (2) it is possible to explain the correlations of the data collected in many runs of a Bell experiment in terms of what happens in each single run, we provided an explicit proof that every theory that accounts for the quantum violation of Bell inequalities must violate parameter independence, thereby being clearly nonlocal. 
An analysis of the violation of outcome independence of quantum mechanics shows that it also is a manifestation of nonlocality.
Indeed, the only case in which this violation is not a direct consequence of a nonlocal process, is that in which it is caused by lack of information about the common past $\pi$.
Nevertheless, a theory taking into account this information must necessarily violate parameter independence.

{To conclude we wish to stress once more that, assuming that the empirical predictions of quantum mechanics are correct, the hypothesis of accountability implies that it is not complete.
Therefore, maintaining the completeness of quantum mechanics is equivalent to maintaining that a phenomenon exists, for which no account is possible.}

\begin{acknowledgments}
We are grateful to Roger Colbeck for useful discussions, to Tomer Barnea, {Christian Beck, Detlef D\"urr,} Dustin Lazarovici, and Samuel Portmann for  helpful comments on an earlier version of this manuscript, and to the organizers of the conferences ``Quantum Malta 2012" and  ``Quantum Theory without Observers III"---both financially supported by the COST action ``Fundamental Problems in Quantum Physics"---during which part of this work was done. 
We acknowledge the financial support of the Elite Network of Bavaria and the Swiss NCCR ``Quantum Science and Technology". 
\end{acknowledgments}

\appendix

\section{Useful proofs}

\subsection{Equivalence of Eq.~\eqref{eq:BellLocality} and Eq.~\eqref{eq:BellLocality2}}
\label{App:BellLocal}

To prove that Eq.~\eqref{eq:BellLocality} implies Eq.~\eqref{eq:BellLocality2}, 
we sum Eq.~\eqref{eq:BellLocality} over all possible values of $X$; the normalization of conditional probability then gives
\begin{equation}
\condpr{Y}{A,B,\lambda}=\condpr{Y}{B,\lambda} .
\end{equation}
Substituting this back into Eq.~\eqref{eq:BellLocality} and using the conditional probability formula~\eqref{eq:Id}  we get
\begin{equation}
\condpr{X}{A,\lambda} =\condpr{X}{Y,A,B,\lambda} ,
\end{equation}
which is the first condition of Eq.~\eqref{eq:BellLocality2}. 
Analogously for the second condition. 

To prove the converse, we substitute Eq.~\eqref{eq:BellLocality2} into Eq.~\eqref{eq:Id} to obtain
\begin{equation}\label{eq:Int1}
\condpr{X,Y}{A,B,\lambda} 
	= \condpr{X}{A,\lambda}\,  \condpr{Y}{A,B,\lambda}   .
\end{equation}

{The conditional probability formula also gives (cf.\ Eq.~\eqref{eq:Id}) 
\begin{equation}
\condpr{X,Y}{A,B,\lambda} 
	= \condpr{Y}{X,A,B,\lambda}\,  \condpr{X}{A,B,\lambda}   ,
\end{equation}
that together with Eq.~\eqref{eq:BellLocality2} becomes
\begin{equation}
\condpr{X,Y}{A,B,\lambda} 
	= \condpr{Y}{B,\lambda}\,  \condpr{X}{A,B,\lambda}   .
\end{equation}
Summing over $X$ and substituting into Eq.~\eqref{eq:Int1} completes the proof.
}

\subsection{Equivalence of Eq.~\eqref{eq:FreeChoice}  and the conjunction of Eqs.~\eqref{eq:NoConspiracy} and~\eqref{eq:ParameterIndependence}}
\label{App:FR}

{Let us begin by proving that Eqs.~\eqref{eq:NoConspiracy} and~\eqref{eq:ParameterIndependence} imply Eq.~\eqref{eq:FreeChoice}. 
From Bayes' theorem we have
\begin{equation}
\condpr{A}{Y,B,\lambda}
= \frac{\condpr{A}{B,\lambda}}{\condpr{Y}{B,\lambda}} \condpr{Y}{A,B,\lambda}  .
\end{equation}
Using Eq.~\eqref{eq:ParameterIndependence} we then get
\begin{equation}
\condpr{A}{Y,B,\lambda} =  \condpr{A}{B,\lambda},
\end{equation}
that together with Eq.~\eqref{eq:NoConspiracy} gives the first condition of Eq.~\eqref{eq:FreeChoice}; the second condition can be proven similarly. 
}

{To show the converse,\footnote{That the FR assumption implies no-conspiracy was already mentioned in Ref.~\cite{ColbeckRenner12a} and that it implies parameter independence  was also proved in Lemma 1 of Ref.~\cite{ColbeckRenner12}.}
consider the conditional probability formula
\begin{align}
\condpr{Y,A}{B,\lambda} &= \condpr{A}{Y,B,\lambda} \condpr{Y}{B,\lambda} 
\\
&= \pr{A} \condpr{Y}{B,\lambda} ,
\end{align}
where the last line follows from Eq.~\eqref{eq:FreeChoice}; summing over $Y$ gives the first line of Eq.~\eqref{eq:NoConspiracy}.
Moreover, by virtue of Bayes' theorem,
\begin{equation}
\condpr{Y}{A,B,\lambda} 
	= \frac{ \condpr{Y}{B,\lambda} }{ \condpr{A}{B,\lambda} } 
		\condpr{A}{Y,B,\lambda}.
\end{equation}
Then, using Eq.~\eqref{eq:FreeChoice} and the first line of Eq.~\eqref{eq:NoConspiracy} we have the second line of Eq.~\eqref{eq:ParameterIndependence}.
}
The complementary set of conditions can be analogously proved.


\begin{thebibliography}{99}

\bibitem{EPR35} A.~Einstein, B.~Podolsky, and N.~Rosen, ``Can quantum-mechanical description of physical reality be considered complete?'' Phys.~Rev.~{\bf47} (10), 777--780 (1935).

\bibitem{Bell1964} J.~S.~Bell, ``On the Einstein-Podolsky-Rosen paradox,'' Physics (NY) {\bf 1} (3), 195-200 (1964).

\bibitem{Aspect:1999} A.~Aspect, `` Bell's inequality test: more ideal than ever,'' Nature {\bf 398} (6724), 189-190 (1999).

\bibitem{GiustinaMechRamelow2013} M.~Giustina, A.~Mech, S.~Ramelow, B.~Wittmann, J.~Kofler, J.~Beyer, A.~Lita, B.~Calkins, T.~Gerrits, Sae~Woo Nam,
R.~Ursin, and A.~Zeilinger., ``Bell violation using entangled photons without the fair-sampling assumption,'' Nature {\bf 497}, 227--230 (2013).


\bibitem{ChristensenMcCuskerAltepeter2013} B.~G. {Christensen}, K.~T. {McCusker}, J.~{Altepeter}, B.~{Calkins}, T.~{Gerrits}, A.~{Lita}, A.~{Miller}, L.~K. {Shalm}, Y.~{Zhang}, S.~W. {Nam}, N.~{Brunner}, C.~C.~W. {Lim}, N.~{Gisin}, and P.~G. {Kwiat}
, ``Detection-loophole-free test of quantum nonlocality, and applications,''
Phys. Rev. Lett. 111, 130406 (2013).

\bibitem{Brunner:2013} N. Brunner, D. Cavalcanti, S. Pironio, V. Scarani, and S. Wehner, ``Bell nonlocality,'' Rev. Mod. Phys. {\bf 86}, 419 (2014).


\bibitem{ClauserHorneShimony1969} J. F. Clauser, M. A. Horne, A. Shimony, and R. A. Holt, ``Proposed experiment to test local hidden-variable theories,'' Phys. Rev.
Lett. {\bf 23}, 880-884 (1969).

\bibitem{ClauserShimony1978} J.~F. Clauser, and A.~Shimony, ``Bell's theorem. Experimental tests and implications,'' Reports on Progress in Physics, {\bf 41} (12), 1881 (1978).

\bibitem{Jarrett1984} J. P. Jarrett, ``On the physical significance of the locality conditions in the Bell arguments,'' No\^us {\bf 18} (4),  569--589 (1984).

\bibitem{Griffiths1987} R.~B. Griffiths. ``Correlations in separated quantum systems: A consistent history analysis of the EPR problem,''  Am. J. Phys. {\bf 55} (1), 11 (1987).

\bibitem{Shimony} A. Shimony, ``An exposition of {Bell}'s theorem,'' published in Sixty-Two Years of Uncertainty, edited by A. Miller (Springer US, 1990), vol. 226 of NATO ASI Series, pp. 33--43

\bibitem{Mermin} N. D. Mermin,  ``What is quantum mechanics trying to tell us?'' Am. J. Phys., {\bf 66} (9), 753--767 (1998).

\bibitem{Dickson1998} W. M. Dickson, {\it Quantum Chance and Non-locality: Probability and Non-locality in the Interpretations of Quantum Mechanics} (Cambridge University Press, Cambridge, 1998).

\bibitem{BertlmannZeilinger2002} R.~A. Bertlmann, and A.~Zeilinger, eds., {\it Quantum (Un)speakables: From Bell to Quantum Information} (Springer, 2002).

\bibitem{Bell2004}  J. S. Bell, {\it  Speakable and Unspeakable in Quantum Mechanics}
(Cambridge University Press, Cambridge, 2004).


\bibitem{Norsen2009} T.~Norsen, ``Local causality and completeness: Bell vs. Jarrett,'' Found. Phys. {\bf 39} (3), 273--294 (2009).

\bibitem{Gisin2012} N.~{Gisin}, ``Non-realism: deep thought or a soft option?
'' Found. Phys. {\bf 42} (1), 80--85 (2012).

\bibitem{SeevinckUffink2011} M. P. Seevinck, and J.~Uffink, ``Not throwing out the baby with the bathwater: Bell's condition of local causality mathematically `sharp and clean',''
in {\it Explanation, Prediction, and Confirmation}, vol.~2, edited by  D.~Dieks, W.~J.~Gonzalez, S.~Hartmann, T.~Uebel, M.~Weber (Springer Netherlands, 2011), pp.~425-450.

\bibitem{CavalcantiWiseman2012} E.~G.~Cavalcanti, and H.~M.~Wiseman, ``Bell nonlocality, signal locality and unpredictability (or What Bohr could have told Einstein at Solvay had he known about Bell experiments),'' Found. Phys. {\bf 42} (10), 1329--1338 (2012).

\bibitem{Henson2013} J. Henson, ``Non-separability does not relieve the problem of Bell's theorem,'' Found. Phys. {\bf43} (8), 1008--1038 (2013).


\bibitem{Berge} B.-G. Englert,  ``On quantum theory'' Eur. Phys. J. D, {\bf 67}, 238 (2013).


\bibitem{ColbeckRenner2011} R. Colbeck and R. Renner, ``No extension of quantum theory can have improved predictive power,'' Nat. Commun. {\bf 2}, 411 (2011).


\bibitem{Norsen2011} T. Norsen, ``John S. Bell's concept of local causality,''Am.~J.~Phys., {\bf 79} (12), 1261--1275 (2011).

\bibitem{Bricmont1999} J. Bricmont, ``What is the meaning of the wave function?'' In {\it  Fundamental interactions: from symmetries to black holes} (1999), pp.~53--67.


\bibitem{Bancal2012} J.-D. Bancal, S. Pironio, A. Ac\'in, Y.-C. Liang, V. Scarani, and
N. Gisin, ``Quantum non-locality based on finite-speed causal influences leads to superluminal signalling,'' Nature Phys. {\bf 8} (12), 867--870 (2012).

\bibitem{Barnea2013} T. J. Barnea, J.-D. Bancal, Y.-C. Liang, and
N. Gisin, ``{A tripartite quantum state violating the hidden influence constraints},'' Phys. Rev. A {\bf 88}, 022123 (2013).

\bibitem{PopescuRohrlich:1994} S. Popescu and D. Rohrlich, ``Quantum nonlocality as an axiom,'' Found. Phys. {\bf 24} (3), 379--385 (1994).

\bibitem{Barrett:2005} J. Barrett, N. Linden, S. Massar, S. Pironio, S. Popescu,
and D. Roberts, ``Nonlocal correlations as an information-theoretic resource,'' Phys. Rev. A {\bf 71} (2), 022101 (2005).

\bibitem{GhirardiRomano2013} G.~{Ghirardi} and R.~{Romano}, ``Comment on 'Is a system's wave function in one-to-one correspondence with its elements of reality?' [arXiv:1111.6597]'', arXiv:1302.1635v1 (2013).

\bibitem{GhirardiRomano2013a} G.~{Ghirardi} and R.~{Romano}, ``About possible extensions of quantum theory'', Found. Phys. {\bf 43}, 881 (2013).

\bibitem{ColbeckRenner2013} R.~{Colbeck} and R.~{Renner},  ``A short note on the concept of free choice'', arXiv:1302.4446v1 (2013).

\bibitem{DewittGraham1973}
B.~S. {Dewitt} and N.~{Graham}, editors,
 {\em {The Many-Worlds Interpretation of Quantum Mechanics}}
 (Princeton University Press, Princeton, New Jersey, 1973).

\bibitem{ColbeckRenner12a} R. Colbeck and R. Renner, ``Is a System?s Wave Function in One-to-One Correspondence with Its Elements of Reality?", Phys. Rev. Lett. { \bf 108}, 150402  (2012).


\bibitem{ColbeckRenner12} R. Colbeck and R. Renner, ``The completeness of quantum theory for predicting measurement outcomes", arXiv:1208.4123 (2012).


\end{thebibliography}
\end{document}